\documentclass[12pt,aps,preprint,nofootinbib]{revtex4}

\usepackage{epsfig}
\usepackage{amssymb}
\usepackage{amsmath}
\usepackage{amsfonts}
\usepackage{graphics}
\usepackage{cancel}
\usepackage{bbm}
\usepackage{mathrsfs}

\newcommand{\Refs}{Refs.}
\newcommand{\Ref}{Ref.}
\newcommand{\Sec}{Sec.}
\newcommand{\Secs}{Secs.}

\newcommand{\Tab}{Tab.}

\newcommand{\eq}{Eq.}

\newcommand{\Fig}{Fig.}

\newcommand{\Co}{\mathcal{C}}
\newcommand{\bld}[1]{\boldsymbol{#1}}
\newcommand{\bea}{\begin{eqnarray}}
\newcommand{\eea}{\end{eqnarray}}

\newcommand{\opbraket}[3]{\left< #1 \left| #2 \right| #3 \right>}

\newcommand{\be}{\begin{equation}}
\newcommand{\ee}{\end{equation}}

\newcommand{\ba}{\begin{array}}
\newcommand{\ea}{\end{array}}

\newcommand{\ie}{\emph{i.e.}}
\newcommand{\eg}{\emph{e.g.}}
\newcommand{\cf}{\emph{c.f.}}

\newcommand{\Slash}[1]{\ooalign{\hfil$ / $\hfil\crcr$#1$}} 

\allowdisplaybreaks

\begin{document}
\title{New physics searches at near detectors of neutrino oscillation experiments}

\author{Stefan Antusch}
\email[]{antusch@mppmu.mpg.de}
\affiliation{Max-Planck-Institut f\"ur Physik
(Werner-Heisenberg-Institut), F\"ohringer Ring 6, 80805 M\"unchen,
Germany}

\author{Mattias Blennow}
\email[]{blennow@mppmu.mpg.de}
\affiliation{Max-Planck-Institut f\"ur Physik
(Werner-Heisenberg-Institut), F\"ohringer Ring 6, 80805 M\"unchen,
Germany}

\author{Enrique Fernandez-Martinez}
\email[]{enfmarti@mppmu.mpg.de}
\affiliation{Max-Planck-Institut f\"ur Physik
(Werner-Heisenberg-Institut), F\"ohringer Ring 6, 80805 M\"unchen,
Germany}

\author{Toshihiko Ota}
\email[]{toshi@mppmu.mpg.de}
\affiliation{Max-Planck-Institut f\"ur Physik
(Werner-Heisenberg-Institut), F\"ohringer Ring 6, 80805 M\"unchen,
Germany}

\begin{abstract}
We systematically investigate the prospects of testing new physics with tau sensitive near detectors at neutrino oscillation facilities. For neutrino beams from pion decay, from the decay of radiative ions, as well as from the decays of muons in a storage ring at a neutrino factory, we discuss which effective operators can lead to new physics effects. Furthermore, we discuss the present bounds on such operators set by other experimental data currently available. For operators with two leptons and two quarks we present the first complete analysis including all relevant operators simultaneously and performing a Markov Chain Monte Carlo fit to the data. 
We find that these effects can induce tau neutrino appearance probabilities as large as $\mathcal O(10^{-4})$, which are within reach of forthcoming experiments.
We highlight to which kind of new physics a tau sensitive near detector would be most sensitive. 

\end{abstract}


\preprint{MPP-2010-51}

\maketitle

\section{Introduction}
\label{sec:intro}

With the expected results of the Atlas and CMS experiments at the LHC, particle physics will enter a new era by directly exploring physics at TeV energies. Complementary to the direct tests, flavour experiments such as LHCb and SuperBelle will search for indirect signs of new physics. On similar time scales, neutrino physics is also entering a new era. Long baseline neutrino oscillation experiments are aiming at a first measurement of the remaining unknown leptonic mixing parameters $\theta_{13}, \delta_{\mathrm{MNS}}$, and the sign of $\Delta m^2_{\mathrm{31}}$, and will determine the other parameters with unprecedented precision. 

For this purpose, neutrino oscillation facilities exploit beams from various sources: currently from pion decay (conventional beams and super-beams~\cite{Itow:2001ee,Ayres:2004js}) and in the future possibly also from the decay of radiative ions ($\beta$-beams~\cite{Zucchelli:2002sa,betabeam}) or from the decays of muons in a storage ring at a neutrino factory~\cite{Geer:1998iz,DeRujula:1998hd,Cervera:2000kp,Bandyopadhyay:2007kx,ids}. In addition to measuring the leptonic parameters, neutrino oscillation facilities can also be excellent probes of new physics, especially in the lepton sector. 

To probe new physics, the near detectors of neutrino oscillation experiments are a powerful tool, in particular when they are capable of detecting tau leptons, as has for example been pointed out in \Refs~\cite{Ota:2002na,Antusch:2006vwa}. With the above mentioned neutrino beams, tau appearance (above a comparatively low background level) would signal new physics. 

Already present beam experiments like MINOS~\cite{Michael:2006rx} could be suitable for performing such a new physics search with a tau sensitive near detector. This idea, known as the Main Injector Non-Standard Interaction Search (MINSIS)~\cite{MINSIS} has recently been discussed, \eg, at \Ref~\cite{MINSIS2009Madrid}. The physics reach of such an experiment  depends, from the theoretical side, on how well the relevant processes at neutrino production and detection are already constrained by other experimental data available. From the experimental side, it is mainly limited by the ability to discriminate a possible signal from the background. 

To explore the potential of testing new physics with near detectors at neutrino oscillation facilities, we systematically investigate the present bounds on higher dimensional operators that can produce tau appearance signals at such experiments.  
In \Sec~\ref{sec:efflag} we introduce the effective operator formalism used to encode the new physics effects that can lead to signals at near detectors and classify them in three types. In \Secs~\ref{sec:quark}, \ref{sec:4L} and~\ref{sec:NU} we discuss the present bounds on each of the types of operators and the extensions of the SM that can lead to the least constrained combination of operators. Finally in \Sec~\ref{sec:conclusions} we summarize the present constraints on each of the processes and the probability level that a near detector would require to probe new physics in each sector.

\section{New physics at near detectors}
\label{sec:efflag}

In general, at a near detector experiment, non-standard neutrino interactions (NSI) can modify both the neutrino production mechanism as well as the neutrino detection.
The NSI do not need to be flavour diagonal, thus leading to striking lepton flavour conversion signals at near detectors, where the standard neutrino oscillations have not developed yet. Indeed, $\nu_\tau$ could be produced in pion or muon decays via new flavour-changing effects, signaling physics beyond the Standard Model (SM) if they are measured at a near detector from such a beam. 
Conversely, neutrino detection via inverse $\beta$ decay (neutrino-nucleon charged current scattering) can be affected and a $\nu_e$ or $\nu_\mu$ can be detected in association with a $\tau$.
It should be noted that, while the new physics we will discuss generally also induce NSI effects in the neutrino flavor propagation in matter, these effects do not have enough time to develop to be measurable in a near detector and we will therefore not discuss them further.

In comparison to disappearance searches, which would be dominated by systematic errors on the neutrino flux normalization and cross sections as well as the response of the detector, the search for neutrino flavor appearance turns out to be an excellent probe of new physics.
Moreover, the bounds on lepton flavour violation in the $e$-$\mu$ sector are generally stronger due to very well constrained processes such as muon to electron conversion in nuclei. In addition, the intrinsic SM backgrounds in the neutrino beams and the detector are much larger for the electron and muon flavours.
Therefore, the prospects to search for new physics are best explored via neutrino
flavour conversion effects to tau neutrinos.

In order to parametrize the new physics that can give rise to a tau appearance signal we shall consider an effective theory approach where the Lagrangian has the form
\begin{equation}
	\mathcal L = \mathcal L_{SM} + \mathcal L_{\rm Weinberg} + \mathcal L_{d=6},
\end{equation}
with $\mathcal L_{SM}$ being the SM Lagrangian, $\mathcal L_{\rm
Weinberg}$ the $d=5$ Weinberg operator giving rise to neutrino
masses~\cite{Weinberg:1979sa}, and $\mathcal L_{d=6}$ a collection of
effective gauge invariant $d=6$ operators that can modify the neutrino
production or detection in a way that leads to a tau signal at a
near detector. There are several such operators~\cite{Buchmuller:1985jz,Bergmann:1998ft,Bergmann:1999pk,Davidson:2001ji,Broncano:2002rw,Broncano:2003fq,Ibarra:2004pe,Abada:2007ux,Biggio:2009nt,Biggio:2009kv}, depending on the production mechanism of the neutrino beam ($\pi$ decay for conventional neutrino beams and super-beam experiments, $\beta$ decay for $\beta$-beam and reactor experiments and $\mu$ decays for neutrino factories). 
We will classify the new physics operators in three different categories:
\begin{itemize}

\item Four-fermion operators involving two leptons and
      two quarks. These operators, when involving a neutrino and a charged lepton, 
      can modify the neutrino production through $\pi$ and $\beta$
      decay, as well as its detection via inverse $\beta$ decay, 
      thus leading to signals at near detectors. 
      We will refer to these operators as $2L2Q$. 

\item Four-fermion operators involving only leptons. These operators
      can modify the neutrino production through muon decays, 
      thus leading to signals at near detectors in a neutrino factory facility. We will refer to these operators as $4L$.

\item ``Kinetic'' operators involving a pair of Higgs doublets and a pair of leptons with a derivative. After the Higgs field develops its vacuum expectation value (vev) these operators contribute to the leptonic kinetic terms and, upon their canonical normalization, induce deviations from unitarity of the leptonic mixing matrix, modifying the $W$ and $Z$ couplings. Such a non-unitary matrix implies flavour conversion effects at zero distance since the flavour eigenstates are no longer orthogonal, thus leading to the desired signals at near detectors. We will refer to these operators as non-unitarity ($NU$) operators.

\end{itemize}  

Here we will study the constraints that the processes related by gauge
invariance to the lepton flavour violating production and detection of
neutrinos imply. Since neutrinos and charged leptons belong to
the same 
$SU(2)$ multiplets, 
lepton flavour violation (LFV)
in the neutrino sector is usually accompanied by related processes in the charged lepton sector. However, it has been shown \cite{Bergmann:1998ft,Bergmann:1999pk,Antusch:2008tz,Gavela:2008ra} that for the $NU$ and $4L$ operators, it is possible to cancel all contributions to processes with four charged fermions while lepton flavour violation is still present in the neutrino sector. For the $2L2Q$ operators, it is trivial to expand the operators in components and see that this cancellation is not possible for the four-charged-fermion (4CF) operators with quarks, since both up and down quarks carry electromagnetic charge and both contributions would need to cancel simultaneously. Therefore, in the study of these operators, care must be taken to include the bounds that are already present on semi-leptonic flavor violating processes. 

In the next sections we will use the bounds on lepton flavour violating processes to derive new bounds on the $2L2Q$ operators and review the existing bounds on the $4L$ and $NU$ operators. We will also discuss the implications for near detector neutrino experiments, detailing the minimum oscillation probability level that would need to be probed in order to improve over the present constraints. We will also discuss briefly which possible new physics models could give rise to the least constrained operators described.

\section{New physics with quarks: $\bld{2L2Q}$ operators}
\label{sec:quark}

\subsection{Bounds on $\bld{2L2Q}$ operators from $\tau$ decays}
\label{sec:decaybounds}
 
In order to derive bounds on the $2L2Q$ operators through flavour changing processes involving four charged fermions, we will use the following basis of gauge invariant operators~\cite{Buchmuller:1985jz} 
\begin{eqnarray}
	(\mathcal O^1_{LQ})_{\alpha}^{\phantom\alpha\beta}
	&=&
	[\overline{L}^\beta \gamma^\rho L_\alpha][\overline{Q}\gamma_\rho Q],
	\\
	(\mathcal O^3_{LQ})_{\alpha}^{\phantom\alpha\beta}
	&=&
	[\overline{L}^\beta \gamma^\rho \tau^a L_\alpha][\overline{Q}\gamma_\rho \tau^a Q],
	\\
	(\mathcal O_{ED})_{\alpha}^{\phantom\alpha\beta}
	&=&
	[\overline{L}^\beta E_\alpha][\overline{D} Q],
	\\
	(\mathcal O_{EU})_{\alpha}^{\phantom\alpha\beta}
	&=&
	[\overline{L}^\beta E_\alpha]i\tau^2[\overline{Q} U]^T,
\end{eqnarray}
where $L_{\alpha}$ is the left-handed $SU(2)$ lepton doublet 
with flavour $\alpha$, $Q$ is the first generation quark doublet 
while $E_{\alpha}$, $D$, and $U$ are the right-handed charged lepton
with flavour $\alpha$, 
down quark, and up quark fields, respectively. 
The matrices $\tau^a$ refer to the generators of the $SU(2)$ symmetry.
Each of these operators $\mathcal O$ will be assumed to appear in
$\mathcal L_{d=6}$ with the corresponding coefficient $2\sqrt{2}G_F
\mathcal C$, \ie, $2\sqrt{2} G_F \mathcal
(\mathcal{C}_{EU})_\beta^{\phantom\beta\alpha}$ as the coefficient of
$(\mathcal O_{EU})_{\alpha}^{\phantom\alpha\beta}$ and so on. 
Here, $G_{F}$ is the Fermi coupling constant.
Since the first two operators are self-conjugate, their coefficients are Hermitian, while the coefficients of the last two operators need not be (however, the conjugate term must also be added to the Lagrangian).

In this paper, we are mainly concerned with the possible signals of new physics at a near tau detector at a neutrino oscillation experiment. The operator coefficients that contribute to these processes are
\[
	(\Co^1_{LQ})_\mu^{\phantom\mu\tau}, \ (\Co^3_{LQ})_\mu^{\phantom\mu\tau}, \ (\Co_{ED})_\mu^{\phantom\mu\tau}, \ (\Co_{EU})_\mu^{\phantom\mu\tau}, \ (\Co_{ED}^\dagger)_\mu^{\phantom\mu\tau}, \ {\rm and} \ 
	(\Co_{EU}^\dagger)_\mu^{\phantom\mu\tau},
\]
for $\nu_\mu$ beams, as well as the corresponding operators with the
muons exchanged for electrons for $\nu_e$ beams, and we will study what bounds can be derived on them with present data. In
order to compute these bounds, we consider the Particle Data Group
(PDG)~\cite{Amsler:2008zzb} bounds on several different lepton flavor
violating decays of the $\tau$ (see \Tab~\ref{tab:operator})\footnote{%
Some of the bounds on the effective operators could be 
improved by $\mu^{\pm} \tau^{\mp}$ signal searches at
hadron colliders in the future.
See \Ref~\cite{Han:2010sa} for a recent study.
}.
Notice that there is some admixture of strange quarks in some of the
mesons involved in the decays, \eg, in the $\eta$ meson. The bounds from such processes
on the coefficients listed above should be correlated with 
the effective interactions with strange quarks.
We therefore also include the possible NSI with strange quarks through the coefficients $\Co_{LQ'}^1$, $\Co_{LQ'}^3$, and $\Co_{ES}$ with the corresponding operators involving the second generation doublet $Q'$ and the right-handed strange singlet $S$. In general, $\Co_{LQ'}^1$ and $\Co_{LQ'}^3$ will only appear in the combination $\Co_{LQ'}^1 + \Co_{LQ'}^3$, which is the coefficient of the operator combination that selects the strange component of the $Q'$ doublet, so that the charmed quarks do not have to be taken into account.

\begin{table}[tb]
{\footnotesize
\begin{tabular}{|l|c|c|c|}
\hline
{\bf Process} 
&{\bf Prefactor}  
& {\bf Relevant combination of coefficients} 
& {\bf BR bound}
\\
\hline \hline
$\tau \rightarrow \ell \rho$
&
$1.7 $ 
& 
$\left| \mathcal{C}_{LQ}^{3} \right|^{2}$
& 
\begin{tabular}{r@{$\cdot$}l}
6.8 & $10^{-8}$ \\[-.2cm]
6.3 & $10^{-8}$
\end{tabular}
\\
\hline
$\tau \rightarrow \ell \omega$
&
$1.4 $ 
& 
$\left| \mathcal{C}_{LQ}^{1} \right|^{2}$
&
\begin{tabular}{r@{$\cdot$}l}
8.9 & $10^{-8}$ \\[-.2cm]
1.1 & $10^{-7}$
\end{tabular}
\\
\hline
$\tau \rightarrow \ell \phi$
&
$0.84 $
&
$
\left|
\mathcal{C}_{LQ'}^{1}
+
\mathcal{C}_{LQ'}^{3}
\right|^{2}
$
&
\begin{tabular}{r@{$\cdot$}l}
1.3 & $10^{-7}$ \\[-.2cm]
7.3 & $10^{-8}$
\end{tabular}
\\
\hline
$\tau \rightarrow \ell \pi$
&
$0.69 $
&
$\left| \mathcal{C}_{LQ}^{3} 
+ \frac{\omega_{\tau}}{2}
\left[
\mathcal{C}_{ED} 
-
\mathcal{C}_{EU}
\right]
\right|^{2}
     +
\frac{\omega_{\tau}^{2}}{4} 
     \bigl| \mathcal{C}_{ED}^{\dagger} 
     -  
     \mathcal{C}_{EU}^{\dagger}
     \bigr|^{2}$
&
\begin{tabular}{r@{$\cdot$}l}
1.1 & $10^{-7}$ \\[-.2cm]
8.0 & $10^{-8}$
\end{tabular}
\\
\hline
$\tau \rightarrow \ell \eta$
&
$0.20 $
&
\begin{minipage}{8cm}
\vspace{0.1cm}
\begin{center}
\begin{eqnarray*}
&&
\left|
\mathscr{F}_{+}
\mathcal{C}_{LQ}^{1}
-
\left[
\mathcal{C}_{LQ'}^{1}
+
\mathcal{C}_{LQ'}^{3}
\right] \phantom{\left\{\frac 12\right\}}\right.
\\&&
 \left.+
\frac{3 m_{\eta}^{2}}{4 m_{s} m_{\tau}}
\mathscr{F}_{-}
\left\{
\frac{1}{2}
\mathscr{F}'
\left[
\mathcal{C}_{EU}
+
\mathcal{C}_{ED}
\right]
-
\mathcal{C}_{ES}
\right\}
\right
|^{2}
\\
&&+
\left(
\frac{3 m_{\eta}^{2}}{4 m_{s} m_{\tau}}
\right)^{2}
\mathscr{F}_{-}^{2}
\Bigl|
\frac{1}{2}
\mathscr{F}'
\left[
\mathcal{C}_{EU}^{\dagger}
+
\mathcal{C}_{ED}^{\dagger}
\right]
-
\mathcal{C}_{ES}^{\dagger}
\Bigr|^{2}
\end{eqnarray*}
\vspace{0.1cm}
\end{center}
\end{minipage}
&
\begin{tabular}{r@{$\cdot$}l}
6.5 & $10^{-8}$ \\[-.2cm]
9.2 & $10^{-8}$
\end{tabular}
\\
\hline
$\tau \rightarrow \ell \pi^{+} \pi^{-}$
&
$0.081 $
&
$\left|
 \mathcal{C}_{ED}
 - 
 \mathcal{C}_{EU}
 \right|^{2}
 +
 \bigl|
\mathcal{C}_{ED}^{\dagger}
 - 
 \mathcal{C}_{EU}^{\dagger}
 \bigr|^{2}$
&
\begin{tabular}{r@{$\cdot$}l}
2.9 & $10^{-7}$ \\[-.2cm]
1.2 & $10^{-7}$
\end{tabular}
\\
\hline
$\tau \rightarrow \ell K^{+} K^{-}$
&
$0.014 $
&
$\left|
 \mathcal{C}_{EU}
 -
 \mathcal{C}_{ES}
 \right|^{2}
 +
 \bigl|
 \mathcal{C}_{EU}^{\dagger}
 - 
 \mathcal{C}_{ES}^{\dagger}
 \bigr|^{2}$
&
\begin{tabular}{r@{$\cdot$}l}
2.5 & $10^{-7}$ \\[-.2cm]
1.4 & $10^{-7}$
\end{tabular}
\\
\hline
$\tau \rightarrow \ell K^{0} \bar{K}^{0}$
&
$0.014 $
&
$\left|
\mathcal{C}_{ED}
 +
 \mathcal{C}_{ES}
 \right|^{2}
 +
 \bigl|
 \mathcal{C}_{ED}^{\dagger}
 + 
\mathcal{C}_{ES}^{\dagger}
 \bigr|^{2}$
&
\begin{tabular}{r@{$\cdot$}l}
3.4 & $10^{-6}$ \\[-.2cm]
2.2 & $10^{-6}$
\end{tabular}
\\
\hline
\end{tabular}
}
\caption{\it Summary of the combination of $2L2Q$ operator coefficients 
with their respective numerical prefactor in the $\tau$ decay branching ratio. In the rightmost column, we quote the PDG bound on the process with the upper value corresponding to $\ell = \mu$ and the lower to $\ell = e$.
Here, the parameters are defined as
$\omega_{\tau} \equiv 
\frac{m_{\pi}}{m_{\tau}} \frac{m_{\pi}}{m_{u} + m_{d}} 
= 1.3$
and 
$\frac{m_{\eta}^{2}}{m_{s} m_{\tau}} = 1.4$.
The ratio of the decay constants are defined as
$\mathscr{F}_{\pm} \equiv 
\frac{F^{8}_{\eta} \pm \sqrt{2} F^{0}_{\eta}}
{F^{8}_{\eta} - \frac{1}{\sqrt{2}} F^{0}_{\eta}}$
and $\mathscr{F}' \equiv 
\frac{F^{8}_{\eta} + 2 \sqrt{2} F^{0}_{\eta}}
{F^{8}_{\eta} - \sqrt{2} F^{0}_{\eta}}$
with $F^{8}_{\eta} = 0.154$ GeV and $F^{0}_{\eta} = 0.025$ 
GeV~\cite{Black:2002wh}, \ie,
$\mathscr{F}_{-} = 0.87$, $\mathscr{F}_{+} = 1.4$, and 
$\mathscr{F}' = 1.9$.}
\label{tab:operator}
\end{table}

In order to compute the branching ratios, we need to compute matrix elements of the type
\begin{equation}
 \mathcal M = \opbraket{\ell \Pi}{\mathcal L_{d=6}}{\tau},
\end{equation}
where $\Pi$ represents the meson (or mesons) involved in a given decay.
For this purpose, we adopt the approach of \Ref~\cite{Black:2002wh}, 
where the authors employed either the Partially Conserved Axial Current (PCAC) 
hypothesis~\cite{Goldberger:1958tr,Adler:1964um,Weisberger:1966ip,Gross:1979ur,Vainshtein:1975sv,Shifman:1975tn}, Vector Meson Dominance (VMD) model~\cite{Sakurai:1960ju,Kroll:1967it,O'Connell:1995wf} or chiral
perturbation theory ($\chi$PT)~\cite{Weinberg:1978kz,Gasser:1983yg,Gasser:1984gg} to the different Lorentz structures of quark currents present,
in order to compute the hadron matrix elements.
Resulting from this procedure are expressions for the partial $\tau$ decay
widths. These expressions consist of a prefactor, which is a function of
measured quantities, such as the SM masses, and
meson decay constants, as well as a combination of the different NSI coefficients relevant to each decay. The resulting decay widths are then compared to the SM decay width of the $\tau$ in order to obtain theoretical predictions for the branching ratios, which are bounded by experiments.
In \Tab~\ref{tab:operator}, we give the relevant operator combinations and prefactors appearing in each decay.
We fully take into account the correlation among 
all the NSI coefficients.

\begin{figure}
\begin{center}
\includegraphics[width=8cm]{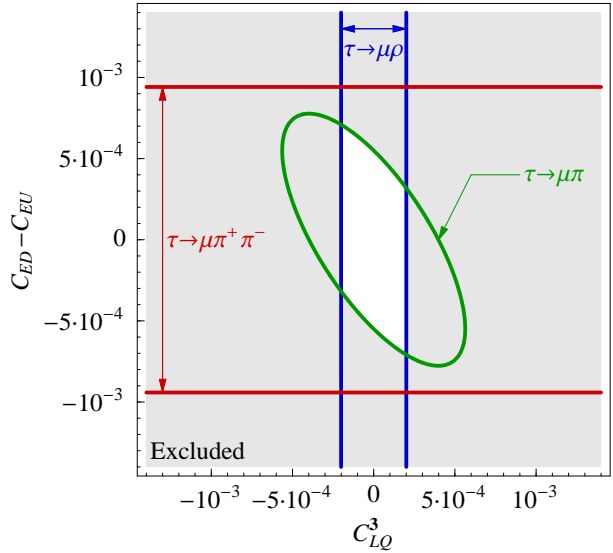}
\caption{\it \label{fig:simpleex} Allowed parameter region from the bounds on the shown $\tau$ decays in the $\Co_{LQ}^3$-$(\Co_{ED}-\Co_{EU})$ plane. It has been assumed that the coefficients are real and that $\Co_{EX} = \Co_{EX}^\dagger$.}
\end{center}
\end{figure}

The general situation, where all coefficients are allowed to appear
independent of each other, is rather involved. Thus, we will first give
an example where we assume that $\Co_{EX} = \Co_{EX}^\dagger$ with $X
\in \{U, D\}$ and that all coefficients are real. In this scenario, the only parameters that appear are $\Co_{LQ}^3$ and the combination $\Co_{ED}-\Co_{EU}$. In \Fig~\ref{fig:simpleex}, we show how different decays constrain this parameter space. Even if very simplified, this figure gives a flavour of the general situation, where the allowed regions are ellipsoids in the parameters on which they depend. The final bounds are then composed by combining the bounds from all considered decays.

For practical purposes, we derive simultaneous bounds on all of the operator coefficients by adopting a Markov Chain Monte Carlo (MCMC) approach, using the MonteCUBES software~\cite{Blennow:2009pk} for the numerical simulations. This approach also allows the study of correlations among the parameters. The PDG bounds on the branching ratios are used as the width of our likelihood function, which we assume to be Gaussian and centred at vanishing branching ratios. We also assume flat priors for both the absolute values and arguments of the complex coefficients, as well as for the values of real coefficients.

\begin{table}
\begin{center}
\begin{tabular}{|c|c|c|c|c|c|c|} \hline 
  Operator   & $(\Co^1_{LQ})_\alpha^{\phantom\mu\tau}$ & $(\Co^3_{LQ})_\alpha^{\phantom\mu\tau}$ & $(\Co_{ED})_\alpha^{\phantom\mu\tau}$ &
$(\Co_{EU})_\alpha^{\phantom\mu\tau}$ & $(\Co_{ED}^\dagger)_\alpha^{\phantom\mu\tau}$ & $(\Co_{EU}^\dagger)_\alpha^{\phantom\mu\tau}$\\ 
\hline \hline
  $\alpha = \mu$ &   $2.1\cdot 10^{-4}$ &     $1.7\cdot 10^{-4}$   &    $7.2\cdot 10^{-4}$    &  $7.2\cdot 10^{-4}$ & $6.2\cdot 10^{-4}$ & $6.2\cdot 10^{-4}$  \\
\hline
  $\alpha = e$ &   $2.4\cdot 10^{-4}$ &     $1.6\cdot 10^{-4}$   &    $6.9\cdot 10^{-4}$    &  $7.1\cdot 10^{-4}$ & $6.0\cdot 10^{-4}$ & $6.1\cdot 10^{-4}$  \\
\hline
\end{tabular}
\caption{\it \label{tab:bounds} Bounds at the 90~\% posterior probability on the operator coefficients from lepton flavor violating decays. The values have been derived through MCMC methods (see text for details).}
\end{center}
\end{table}

\begin{figure}[t!]
\vspace{-0.5cm}
\begin{center}
\begin{tabular}{cc}
\hspace{-0.55cm} \epsfxsize7.5cm\epsffile{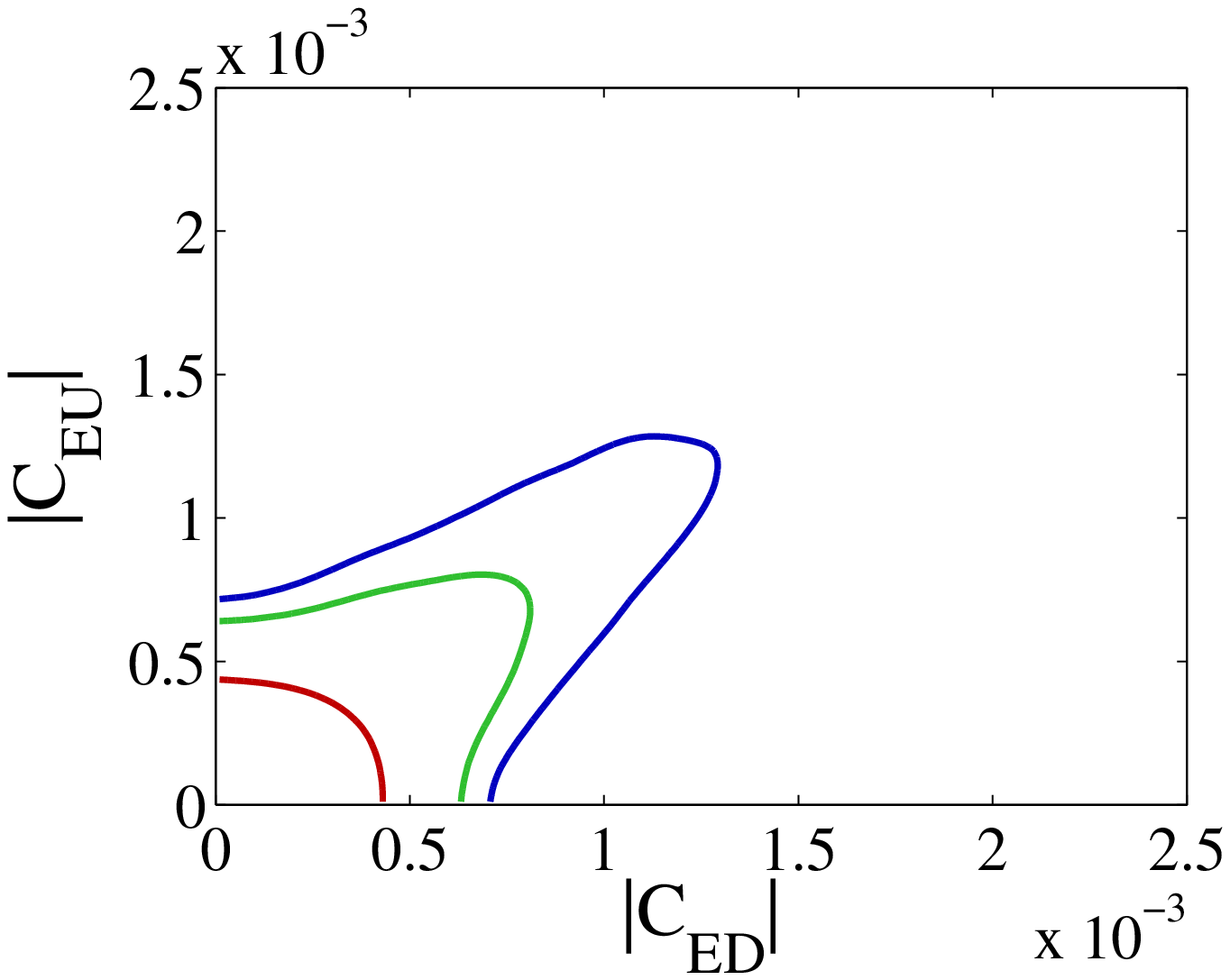} & 
                 \epsfxsize7.5cm\epsffile{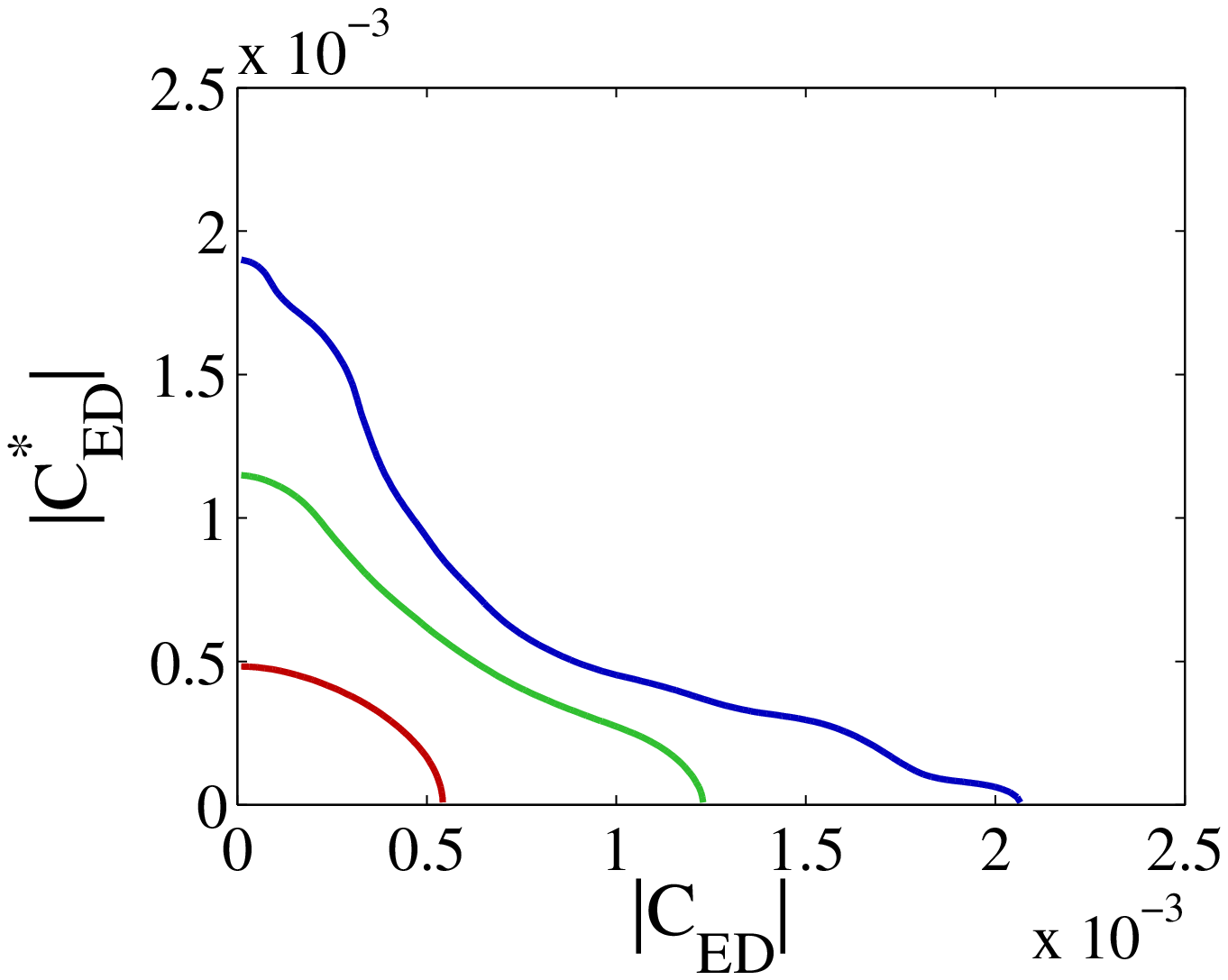} 
\end{tabular}
\caption{\it \label{fig:correlation} The $68$, $90$ and 95~\% posterior probability contours
 as an example of the correlations found among the parameters
 ${(\Co_{ED})_{\mu}}^{\tau}$, 
 ${(\Co_{EU})_{\mu}}^{\tau}$ and 
 ${(\Co_{ED}^\dagger)_{\mu}}^{\tau}$, 
see text for details.} 
\end{center}
\end{figure}

The derived bounds on the operator coefficients $\Co$ after consideration
of the full correlation among them are given in \Tab~\ref{tab:bounds}. 
The bounds on most of the parameters are essentially independent. However, there is a correlation between $\Co_{ED}$ and $\Co_{EU}$, since in most branching ratios they appear in the $\Co_{ED}-\Co_{EU}$ combination, as can be seen in \Tab~\ref{tab:operator}. This direction is thus more strongly bound as can be seen in the left panel of \Fig~\ref{fig:correlation}. A similar correlation is present for $\Co_{ED}^\dagger$ and $\Co_{EU}^\dagger$. Conversely, since in the branching ratios  $\Co_{ED}$ and $\Co_{EU}$ tend to appear in the same combination as $\Co_{ED}^\dagger$ and $\Co_{EU}^\dagger$, respectively, but in an incoherent sum, we find an anticorrelation between $\Co_{ED}$ or $\Co_{EU}$ and $\Co_{ED}^\dagger$ or $\Co_{EU}^\dagger$, since when one is small the present bounds allow the other to be larger. This situation is depicted for $\Co_{ED}$ and $\Co_{ED}^\dagger$ in the right panel of \Fig~\ref{fig:correlation}.

\subsection{Implications for near detectors}
\label{sec:neardetect}

With the bounds on the effective operators derived 
in the former section,
we will now discuss the expected number of signal events 
in a near detector experiment.
Let us first focus on the case of 
a conventional $\nu_{\mu}$ beam from $\pi$ decay. 
The signal process is
\begin{align}
\pi^{+} \rightarrow \mu^{+} &\nu
 \nonumber 
 \\
 & \nu N \rightarrow \tau^{-} X, \nonumber
\end{align}
where $N$ is a target nucleon in a detector 
and $X$ represents all the final states of 
the neutrino-nucleon scattering.
The $\mu$-$\tau$ flavour change can enter at the neutrino beam source
or in the detection process~\cite{Grossman:1995wx,GonzalezGarcia:2001mp}.
The rate of this process is calculated with 
a coherent sum of the amplitudes\footnote{%
There is a distance $L$ between the neutrino beam source and the near
detector in a real experiment,
and the contribution from standard neutrino oscillation 
for $\nu_{\mu} \rightarrow \nu_{\tau}$ 
may not be negligible.
The effect can be taken into account by coherently adding 
the oscillation amplitude $\langle 
 \tau^{-} X | 
 \mathcal{L}_{\rm CC} 
 | \nu_{\tau} N 
 \rangle
 \langle \nu_{\tau} | {\rm e}^{-{\rm i} H L} | \nu_{\mu} \rangle
 \langle 
 \nu_{\mu} \mu^{+}|
 \mathcal{L}_{\rm CC}
 | \pi^{+} 
 \rangle
 $ into 
Eq.~\eqref{eq:Mtotal-pi}.
Here, $H$ is the neutrino propagation Hamiltonian. 
},
\begin{align}
\mathcal{M}_{\pi\text{-total}}
=
 \langle 
 \tau^{-} X | 
 \mathcal{L}_{\rm CC} 
 | \nu_{\tau} N 
 \rangle
 \langle 
 \nu_{\tau} \mu^{+}|
 \mathcal{L}_{d=6} 
 | \pi^{+} 
 \rangle
 +
 \langle 
 \tau^{-} X | 
 \mathcal{L}_{d=6}  
 | \nu_{\mu} N 
 \rangle
 \langle 
 \nu_{\mu} \mu^{+}|
 \mathcal{L}_{\rm CC}
 | \pi^{+} 
 \rangle,
\label{eq:Mtotal-pi}
\end{align}
where the Lagrangian $\mathcal{L}_{\rm CC}$ is the 
charged current four-Fermi interaction 
induced by $W$ boson exchange.
The first term corresponds to the amplitude with the NSI at the source,
and the second term to that at detection.
Since the size of these NSI amplitudes is characterized
by the coefficients $\mathcal{C}$, which are constrained to 
$\mathcal{O}(10^{-4})$ (see \Tab~\ref{tab:bounds}), 
the ratio $\mathcal{R}$ 
between the $\tau^{-}$ signal events and 
the lepton flavour conserving $\mu^{-}$ events 
induced by the SM interactions is naively expected to be 
of an order of the square of $\mathcal{C}$, \ie, at most $\mathcal{O}(10^{-7}-10^{-8})$.
However, there is an enhancement mechanism of the NSI amplitude 
with a pseudo-scalar quark current in pion decays~(see, \eg, 
\Ref~\cite{Herczeg:1995kd}). This is due to spin conservation, since such an operator would involve fermions of the appropriate chirality
and therefore not require a mass-suppressed chirality flip as in the SM. 
Indeed, the LFV pion decay rate would be given by
\begin{align}
\Gamma( \pi^{+} \rightarrow \nu_{\tau} \mu^{+})
=
\left|
2 {(\mathcal{C}_{LQ}^{3})_{\mu}}^{\tau}
+
\omega_{\mu}
\left[
{(\mathcal{C}_{ED}^{\dagger})_{\mu}}^{\tau}
 -
{(\mathcal{C}_{EU}^{\dagger})_{\mu}}^{\tau}
\right]
\right|^{2}
\Gamma (\pi^{+} \rightarrow \nu_{\mu} \mu^{+}),
\end{align} 
with the chiral enhancement factor $\omega_{\mu}$ defined as\footnote{%
The factor strongly depends on the value of the quark masses.
Here we adopt the PDG average of $m_{u}+m_{d} = 8.0$ MeV~\cite{Amsler:2008zzb}.}
\begin{align}
\omega_{\mu}
=
\frac{m_{\pi}}{m_{\mu}}
\frac{m_{\pi}}{m_{u} + m_{d}}
\simeq 21,
\end{align}
as derived from the PCAC~\cite{Donoghue:1978cj}. From the MCMC global fit we obtain the bound 
\begin{align}
\left|
{(\mathcal{C}_{ED}^{\dagger})_{\mu}}^{\tau}
 -
{(\mathcal{C}_{EU}^{\dagger})_{\mu}}^{\tau} 
\right| < 4.2 \cdot 10^{-4},
\end{align}
for the relevant combination of coefficients.
This translates into a constraint on the signal rate of  
\begin{align}
\mathcal{R} < 7.9 \cdot 10^{-5}.
\end{align}
The current direct bound on this process is given by the short baseline
neutrino oscillation experiments NOMAD~\cite{Astier:2001yj} and 
CHORUS~\cite{Eskut:2007rn}:
\begin{eqnarray}
\mathcal{R}
&<& 1.63 \cdot 10^{-4} \text{ (NOMAD)},
\\
&<& 2.2 \cdot 10^{-4} \text{ (CHORUS)}.
\end{eqnarray}
Recently, a new experimental proposal, MINSIS, for a tau near detector in the NUMI beam at Fermilab has been proposed~\cite{MINSIS,MINSIS2009Madrid}.
The expected sensitivity to $\mathcal{R}$
is estimated to be $\mathcal{O}(10^{-6})$ 
or even better.
The sensitivity to the source and detection NSIs 
in the on-going and forthcoming conventional beam experiments 
with explicit implementation of near detectors 
was discussed 
in Ref.~\cite{Kopp:2007ne}. It
shows that the sensitivity is limited 
with a near detector which does not have capability of tau-detection.

The $2L2Q$ interactions would also affect 
the source and detection processes in a future $\beta$-beam facility,
where pure $\bar{\nu}_{e}$ (or $\nu_{e}$) are
produced through the decays of radioactive ions.
With $e$-$\tau$ flavour violating NSIs,
the produced beam would be contaminated by $\bar{\nu}_{\tau}$.
Thus, the process $\bar{\nu}_{e} N \rightarrow \tau^{+} X$ 
would signal these new physics at a near detector.   
The amplitude for the $\bar{\nu}_{\tau}$ beam production 
can be written as
\begin{eqnarray}
\mathcal{M}_{\beta\text{-source}}
&=&
\langle \bar{\nu}_{\tau} e^{+} N' | \mathcal{L}_{d=6} | N \rangle
\nonumber 
\\
&=&
2 \sqrt{2} G_{F} 
\langle 1 \rangle 
\left\{
{(\mathcal{C}_{LQ}^{3})_{e}}^{\tau}
[\bar{u}(e) \gamma^{0} {\rm P}_{L} v(\nu_{\tau})]
+
\left[
{(\mathcal{C}_{ED}^{\dagger})_{e}}^{\tau}
+
{(\mathcal{C}_{EU}^{\dagger})_{e}}^{\tau}
\right]
[\bar{u} (e) {\rm P}_{L} v(\nu_{\tau})]
\right\}
\nonumber 
\\
&&
-
2 \sqrt{2} G_{F} 
\langle \sigma_{i} \rangle 
{(\mathcal{C}_{LQ}^{3})_{e}}^{\tau}
[\bar{u}(e) \gamma^{i} {\rm P}_{L} v(\nu_{\tau})],
\end{eqnarray} 
in the non-relativistic limit for the nucleon part.
Here, $\langle 1 \rangle$ and $\langle \sigma_{i} \rangle$
are the nuclear matrix elements for the Fermi and 
the Gamow-Teller transitions, respectively.
Contrary to the case of conventional beams, 
there is no enhancement mechanism in the amplitude.
Similarly, no enhancements of the NSI effects would be present 
in the detection process,
which, as for the conventional neutrino beam, would occur through inverse 
$\beta$ decay.
It therefore follows that 
the $2L2Q$ search at a near detector of a beta beam facility  
would be a demanding task given the smallness of the signal.
The sensitivity reach at the beta beam experiment with 
a particular setup of a near detector was studied in 
\Ref~\cite{Agarwalla:2006ht}.
The flux in reactor experiments also originates from 
a beta decay and its sensitivity to the NSIs was discussed 
in \Ref~\cite{Ohlsson:2008gx}. 

\subsection{Connection to models at high energy scale}
\label{sec:models}

According to the discussion in \Sec~\ref{sec:neardetect},
we conclude that
the most relevant $2L2Q$ operators for near detector searches
are $\mathcal{O}_{ED}$ and $\mathcal{O}_{EU}$.
In the following we will discuss which possible
high-energy physics could give rise to these operators in their low-energy effective theories.

The operator $\mathcal{O}_{ED}$ can be decomposed into the following fundamental
interactions through mediator fields~\cite{Bergmann:1998ft,Bergmann:1999pk,Davidson:2001ji}:
\begin{align}
{(\mathcal{O}_{ED})_{\alpha}}^{\beta}
=
\begin{cases}
[\overline{L}^{\beta i} E_{\alpha} 
]
[\overline{D} Q_{i} 
]
,
\text{ with a colour singlet scalar $\Phi ({\bf 2}^{s}_{+1/2})$}, 
\\
-\frac{1}{2}
[\overline{L}^{\beta i} \gamma^{\rho} D^{c} 
]
[\overline{Q^{c}}_{i} \gamma_{\rho} E_{\alpha} 
]
,
\text{ with a colour triplet $(\bar{\bf 3})$ 
 vector $V_{2}({\bf 2}^{v}_{+5/6})$},
\\ 
-\frac{1}{2}
[\overline{L}^{\beta i} \gamma^{\rho} Q_{i} 
]
[\overline{D} \gamma_{\rho} E_{\alpha} 
]
,
\text{ with a colour triplet $({\bf 3})$ 
vector $U_{1}({\bf 1}^{v}_{+2/3})$,}
\end{cases}
\nonumber
\end{align}
where the symbol ${\bf X}^{Z}_{Y}$ for mediator fields 
indicates
their $SU(2)_{L}$ representation ${\bf X}$,
$U(1)_{Y}$ hypercharge $Y$, and 
Lorentz nature $Z \in \{ s, v\}$ 
with $s$ for a scalar and $v$ for a vector. The situation for $\mathcal O_{EU}$ presents a similar structure.

The scalar mediator $\Phi$ has the same charges as the standard 
Higgs doublet. Thus, in a general two Higgs doublet model (THDM),
the additional Higgs doublets could play 
the role of $\Phi$ and mediate the NSI and LFV interactions.
Note, however, that the doublet may be an ``inert'' Higgs field which 
does not get a vacuum expectation value. The other two mediators
are typically referred to as ``leptoquarks''.   

In an explicit model to realise the 2L2Q operators, possible
additional constraints would have to be taken into account and it
is an interesting question whether the model-independent 
bounds derived in this study can be saturated in such a model.  

Similar types of models, and their phenomenological constraints, 
have been studied in the literature. For example, there are studies 
on LFV in the THDM~\cite{Paradisi:2005tk,Kanemura:2005hr,Davidson:2010xv}, 
Higgs mediated LFV in the minimal supersymmetric standard model~\cite{Paradisi:2005tk,Babu:2002et,Dedes:2002rh,Brignole:2004ah,Arganda:2008jj,Herrero:2009tm},
embedding of the doublet scalar mediator into a R-parity violating supersymmetric model~\cite{Barbier:2004ez}, and models with leptoquarks~\cite{Buchmuller:1986zs}.
%

\section{New physics with leptons: $\bld{4L}$ operators}
\label{sec:4L}

\subsection{Bounds on $\bld{4L}$}
\label{sec:bounds4l}

As discussed in \Sec~\ref{sec:efflag}, in the case of $4L$ interactions, a $d=6$ operator exists such that 
neutrino production and propagation processes can be affected in a flavour violating manner, while the corresponding charged lepton flavour violation is not induced. 
As a result, processes such as the LFV $\tau$ decays studied in \Sec~\ref{sec:decaybounds} cannot be exploited to derive bounds on this other kind of NSI.
Nevertheless, other constraints can be derived.

In the case of $4L$ interactions, there is a $d=6$ operator that avoids 4CF flavour violating processes~\cite{Bergmann:1998ft,Bergmann:1999pk}:
\begin{equation}
{(\mathcal{O}_{LL})_{\alpha}}^{\beta}
=
2
(\overline{L^{c}}_{\alpha} i \tau^{2} L_{\mu})
(\overline{L}^{e} i \tau^{2} L^{c \beta}).
\label{eq:leptonnsi}
\end{equation}
As such, constraining this operator in a model-independent way can be challenging. In particular, its most relevant effect is to modify the
$\mu$ decay. This process is experimentally measured with exquisite precision and used to extract the value of the Fermi constant $G_F$. 
Indeed, the measured value of $G_F$ through $\mu$ decays would be modified in presence of the operators of Eq.~(\ref{eq:leptonnsi}) with coefficients ${(\mathcal{C}_{LL})_{\alpha}}^{\beta}$ as:
\begin{equation}
G_\mu = G_F \left( \left| 1 + {(\mathcal{C}_{LL})_{e}}^{\mu}  \right|^2 + \left| {(\mathcal{C}_{LL})_{e}}^{\tau}  \right|^2 + \left| {(\mathcal{C}_{LL})_{\tau}}^{\mu}  \right|^2 + \left| {(\mathcal{C}_{LL})_{\tau}}^{\tau}  \right|^2\right).
\end{equation}
Hence, in order to derive bounds 
on the coefficients of these operators, the value of $G_F$ should be compared with other process from which it can be extracted but are not similarly affected.
Such an alternative is offered by the quark sector through $\beta$ and
$K$ decays. These decays are used to measure the values of the 
Cabibbo-Kobayashi-Maskawa (CKM) matrix elements $V_{ud}$ and $V_{us}$
and used to test the unitarity relation, experimentally \cite{Amsler:2008zzb}: 
\begin{equation}
|V_{ud}|^2 + |V_{us}|^2 = 1.000 \pm 0.001
\end{equation}
at 90~\% CL. In order to make these measurements, $G_\mu$ is used as a
measurement of the Fermi constant. Thus, barring cancellations between
the first order of ${(\mathcal{C}_{LL})_{e}}^{\mu}$ and the second order
of the rest of the coefficients that violate lepton flavour, the latter
can be bounded to be $|{(\mathcal{C}_{LL})_{\alpha}}^{\beta}| < 0.032$.
For the signal ratio $\mathcal{R}$, this bound implies 
\begin{align}
\mathcal{R} < 1.0 \cdot 10^{-3} \text{ ($G_{F}$ measurement)}.
\end{align}  

\subsection{Connection to models at high energy scale}
\label{sec:models4l}

In the case of leptonic NSI, the operator of \eq~\eqref{eq:leptonnsi} can be mediated by a charged scalar singlet
${\bf 1}^{s}_{-1}$~\cite{Cuypers:1996ia},
which is often contained in models for radiative 
neutrino mass generation~\cite{Zee:1980ai,Zee:1985rj,Babu:1988ki,Ma:1998dn}. 
In this simplest realization, only three couplings of the scalar to the lepton doublets are different from zero and stronger constraints than the ones described in \Sec~\ref{sec:bounds4l} apply. These constraints were studied in \Refs~\cite{Cuypers:1996ia,Antusch:2008tz} and are summarized in \Tab~\ref{tab:summary} along with the weaker model-independent ones.
The operator can in principle also be realized by a combination of the other 
types of mediator fields,
\eg, 
$Z'({\bf 1}^{v}_{0})$ and $W'({\bf 3}^{v}_{0})$, 
in which the couplings between leptons and the mediation fields
must follow a specific relation to cancel 4CF~\cite{Gavela:2008ra}. 

\section{New physics with kinetic operators: $\bld{NU}$}
\label{sec:NU}

\subsection{Bounds on $\bld{NU}$}
\label{sec:boundsnu}

As for the kinetic operators, the effective $d=6$ operator
with Higgs doublets $\phi = (\phi^{+}, \phi^{0})^{T}$
\begin{align}
{(\mathcal{O}_{\text{MUV}})_{\alpha}}^{\beta}
=
(\overline{L}^{\beta} i \tau^{2} \phi^{*})
(i \Slash{\partial}) (\phi^{T} i \tau^{2} L_{\alpha})
\label{eq:OMUV}
\end{align}
leads to non-canonical kinetic terms for neutrinos
after electroweak symmetry breaking and, after their canonical normalization, induces non-unitarity in the  
lepton mixing matrix~\cite{DeGouvea:2001mz,Broncano:2002rw,Broncano:2003fq}.
Through this particular operator non-unitarity is induced in such a way that only neutrino interactions are affected and the rest of the SM Lagrangian remains unchanged. For this reason, it is often called Minimal Unitarity Violation~(MUV). Since the couplings to the SM gauge bosons themselves are modified, they mediate NSI that affect the neutrino production, detection and propagation in any experiment in the same way,
although the gauge interactions for charged leptons, 
which are measured with a high precision and consistent 
with the SM, are not affected.
Thus, the bounds from processes such as the ones studied \Sec~\ref{sec:decaybounds} cannot be applied 
to $NU$ operators, since they
originate from the $SU(2)_{L}$ breaking with the vev of Higgs doublet.
Bounds on the NSI of this form can be found in the
literature~\cite{Langacker:1988up,Langacker:1988ur,Nardi:1994iv,Tommasini:1995ii,Antusch:2006vwa,Antusch:2008tz},  
and are also summarized in \Tab~\ref{tab:summary}.

\subsection{Connection to models at high energy scale}
\label{sec:modelsnu}

The operator in \eq~\eqref{eq:OMUV}
is a typical imprint of the mixing between neutrinos and
heavy SM singlet fermions~\cite{Langacker:1988up,Langacker:1988ur}, such as 
right-handed neutrinos in TeV scale seesaw mechanism (see, \eg, Refs.~\cite{Mohapatra:1986bd,Branco:1988ex})
or Kaluza-Klein modes of bulk right-handed neutrinos
in extra dimensional models~\cite{DeGouvea:2001mz}.
Other attempts to exploit the EWSB in NSI 
were also made with a dimension eight operator with Higgs doublets~\cite{Berezhiani:2001rs,Davidson:2003ha,Antusch:2008tz,Gavela:2008ra}.
However, the construction of a concrete model with the dimension eight operators
is either rather contrived or reduces to the construction at $d=6$ discussed earlier.


\section{Summary and discussion}
\label{sec:conclusions}

%
\begin{table}
\begin{center}
\begin{tabular}{|c|c|c|c|} \hline 
 Beam (channel)  & $2L2Q$ & $4L$ & $NU$ \\
\hline \hline
  $\pi$ ($\mu \rightarrow \tau$)   &   $7.9 \cdot 10^{-5}$ &   n/a   & $4.4 \cdot 10^{-6}$ \\
\hline
  $\beta$ ($e \rightarrow \tau$)   &  $< 10^{-6}$ &   n/a   & $1.0 \cdot 10^{-5}$ \\
\hline
  $\mu$ ($\mu \rightarrow \tau$)   &  $< 10^{-6}$ &   $1.0 \cdot 10^{-3}$ $(3.2 \cdot 10^{-5})$   &   $4.4 \cdot 10^{-6}$ \\
\hline
  $\mu$ ($e \rightarrow \tau$)   &   $< 10^{-6}$ &   $1.0 \cdot 10^{-3}$ $(3.2 \cdot 10^{-5})$   &   $1.0 \cdot 10^{-5}$ \\
\hline
\end{tabular}
\caption{\it \label{tab:summary} Bounds at the 90~\%~CL on the probability
 of tau appearance at a near detector in a neutrino beam
 from $\beta$ decay, $\pi$ decay  or $\mu$ decays for the three types of
 new physics.
Two different values, the bound to the effective four Fermi interactions 
and that with the assumption of the singlet scalar decomposition 
(in parenthesis), 
are shown in the column of leptonic NSI.}
\end{center}
\end{table}

Tau appearance at near detectors of neutrino oscillation facilities provides a very promising window to new physics. In this paper we have systematically investigated which types of new physics can give rise to such a signal and discussed the bounds on the corresponding effective operators from the experimental data currently available. These bounds are summarised in \Tab~\ref{tab:summary} for neutrino beams from pion decay (conventional neutrino beams and super-beams), from the decay of radiative ions ($\beta$-beams) as well as from the decays of muons in a storage ring at a neutrino factory and for the three different types of new physics specified in \Sec~\ref{sec:efflag}. The resulting bounds on the tau appearance probability have to be compared to the estimated background level which has an intrinsic value of $\mathcal O(10^{-6})$ with presently discussed $\tau$ detection technology~\cite{Autiero:2003fu}.

Regarding new physics effects leading to tau appearance in beams from $\pi$ decay, only the operators with two quarks and two leptons ($2L2Q$) and kinetic operators leading to non-unitarity ($NU$) of the leptonic mixing matrix are relevant. While bounds on the $NU$ operators have been discussed already in \Refs~\cite{Langacker:1988up,Langacker:1988ur,Nardi:1994iv,Tommasini:1995ii,Antusch:2006vwa,Antusch:2008tz} and have here only been reviewed, 
for operators with two leptons and two quarks ($2L2Q$ operators) we have presented the first complete analysis including all relevant operators simultaneously and performing a Markov Chain Monte Carlo fit to the data. Taking the chiral enhancement of certain operators in $\pi$ decay into account, we found that current bounds only restrict the tau appearance probability to be smaller than $7.9 \cdot 10^{-5}$, which means that an experiment like MINSIS would be about two orders of magnitude more sensitive to new physics than existing experiments. We have also discussed how extensions of the SM with leptoquarks or additional Higgs doublets could give rise to such $2L2Q$ operators. On the other hand, present bounds on the $NU$ operators restrict the tau appearance probability to be smaller than $4.4 \cdot 10^{-6}$, so that observing this signal above the background level becomes much challenging.

For neutrino beams from $\beta$ and $\mu$ decay, the chiral enhancement is not present and therefore the effects from $2L2Q$ are already too constrained from the current experimental data to lead to a signal above the background level. On the other hand, for tau appearance caused by $NU$ effects, the bounds on the appearance probability are at the level of $10^{-5}$ and future neutrino oscillation facilities might be significantly more sensitive~\cite{FernandezMartinez:2007ms,Goswami:2008mi,Antusch:2009pm,Meloni:2009cg}. The bounds on effects from $NU$ operators are only slightly stronger for facilities with beams from $\pi$ decays so some improvement can be achieved there as well. Finally, we would like to remark that, without including specific new physics generating these operators, there are operators with four leptons ($4L$) which could cause tau appearance probabilities at the level of $10^{-3}$. However, when the operator is generated in an explicit model of new physics (\ie, by a singly charged scalar field), then much stronger bounds apply~\cite{Cuypers:1996ia,Antusch:2008tz}. 

In summary, we found that, although bounds from the current experimental data are already quite strong for some possible new physics effects, there are some types (\cf~\Tab~\ref{tab:summary}) of new physics operators to which tau appearance searches, already envisioned at present neutrino oscillation facilities, are very sensitive. A tau sensitive near detector might be able to probe new physics at sensitivity levels up to two orders of magnitude better than current bounds.

\begin{acknowledgments}

It is a pleasure to acknowledge very interesting discussions, which motivated this work, with Adam Para and with the participants of the workshop organized by Belen Gavela
in Madrid around the MINSIS idea. The work was also encouraged by discussions with Yoshitaka Kuno and Shinya Kanemura. We also want to acknowledge discussions with 
Marc Sher and Deirdre Black on the computation of the hadron matrix elements employed here.
This work was supported by the European Community through the European Commission Marie Curie Actions Framework Programme 7 Intra-European Fellowship: Neutrino Evolution [M.B.] and by the DFG cluster of excellence ``Origin and Structure of the Universe''.

\end{acknowledgments}


\end{document}